\documentclass[twocolumn,showpacs,preprintnumbers,amsmath,amssymb]{revtex4}

\usepackage{graphicx}
\usepackage{dcolumn}
\usepackage{bm}

\begin{document}
\def\v{\vspace{0.4cm}}
\def\n{\noindent}
\def\be{\begin{equation}}
\def\ee{\end{equation}}
\def\bbb{\begin{eqnarray}}
\def\eee{\end{eqnarray}}
\def\bea*{\begin{eqnarray*}}
\def\eea*{\end{eqnarray*}}
\def\ri{{\rm i}}
\def\u{\underline}

\preprint{APS/123-QED}

\title{Stationary Phase and the Theory of Measurement\protect\\
--- $1/N$ expansion ---}

\author{R. Fukuda}
\affiliation{Department of Physics, Faculty of Science and Technology,
 Keio University, Yokohama, Japan}
%
 \email{fukuda@rk.phys.keio.ac.jp}
%

\date{\today}

\begin{abstract}
The measuring process is studied,
 where a macroscopic number $N$ of particles in the detector 
interact with the object.
When $N\rightarrow\infty$, the fluctuation of the object between 
different eigen-states of the operator $O$ to be measured is suppressed, 
frozen to one and the same state while the detector is on.
 During this period, 
 the stationary phase accompanying the macrovariable is established to have a 
one to one correspondence with the eigen-value of $O$.  
 A model is studied  
 which produces the ideal result when
 $N\rightarrow\infty$ and the correction terms are calculated in powers of $1/N$. 
It is identical to the expansion including the fluctuation of the object successively.   
\end{abstract}

\maketitle

\section{\label{intro}Introduction}

The observational problem in quantum mechanics has a long history of debates 
 \cite{referen,DEspagnat,Bell,Genovese,Ghose,Afriat,Gellmann,Mensky}. 
In particular, the role of the docoherence 
due to environmental influence has been widely discussed
 \cite{Wheeler,Zurekk,Omnes,Joos,Schlosshauer}.   
The dynamical reduction model has actually been 
constructed and many recent researches
 are focused on this subject \cite{Pearl,Rimini,Leggett,Caves1,Bassi}. 
Irrespectively of the mechanism that leads to the reduction, 
we need the statistical treatment of the data,
 which is done by using the wave function following the 
rule of the ordinary quantum mechanics. 
 When we apply the quantum mechanics to the detector system,
 composed of a number of microscopic particles,  
two requirements have to be fulfilled;
(1)~any detector variable $X(t)$, the pointer position for example,
should show a non-fluctuating behavior of classical type 
as a function of the time. 
(2)~
 different eigen-states of object operator $O$ to be measured 
have to be mapped onto 
 different values of $X(t)$.
\\
~~We stress here that both (1) and (2) are realized by the 
stationary phase accompanying any macroscopic variables \cite{fukuda} 
 (The precise definition of the macrovariable 
is given in Sec.\ref{macro}). 
 Consider (1) in the path-integral form \cite{Feynman}. Out of many 
fluctuating paths,  
the stationary phase  
selects one particular smooth path denoted as $X^{\rm st}(t)$.
 Then the absolute square of 
the wave function of $X$ integrated by other degrees
 has a non-diffusive peak, equal to 
 the density of a 
classical point-like particle $\delta(X-X^{\rm st}(t))$.  
This is what we observe as a macroscopic object in the ordinary life. 
As for (2), we recall that 
 the measuring device is prepared 
in such a way that $O$ 
interacts, directly or indirectly, with a large number $N$ 
 of particles in the detector. So the Hamiltonian $H_I$
 describing such an interaction 
 may be a function of $O$, coordinates $x_i$'s and the momenta $p_i$'s of all these particles and is $O(N)$. 
Since the Hamiltonian of the object $H_O$ itself is $O(1)$,
 it can be neglected compared with $H_I$. 
Then, for $N\rightarrow\infty$, 
 {\it each eigen-state of $O$ is frozen in the same state
 as long as the detector is on}. Moreover, {\it precisely during this period,
  the detector variable $X(t)$ changes    
 its value depending on the eigen-values of $O$.} 

The desired mapping is realized in this way.
 Note that if 
the object interacts with a finite number of particles, 
the process is simply a quantum mechanical one, nothing to do with the 
measurement.
 
The above observation suggests the $1/N$ expansion scheme for large $N$, which 
 incoorporates successively the
 fluctuation of the object connecting different eigen-states of $O$. 
It is the purpose of this paper to show that this is indeed the case 
by adopting a simple model of separable $X$. It is solvable by $1/N$ expansion
 and we calculate several lower order 
  terms explicitly. 
 The stationary phase was applied to
 the macrovariable in \cite{fukuda} in  
the lowest order but above observations were lacking. 

In Sec.\ref{stationarity}, crucial 
points of the stationary phase are presented
 on which subsequent discussions are based.  
These are not stated in \cite{fukuda}. 
Following Sec.\ref{stationarity}, a
 model is defined in Sec.\ref{modell}, and 
the signal function is introduced, which agrees with the density of the 
classical point-like particle for 
 $N\rightarrow\infty$.
Higher order corrections are studied in Sec.\ref{higher}. 
They are given by the expansion in (fractional) powers of 
$1/N$ and the result is summarized in (\ref{AbBb}) below. 
This type of calculation is surely required since in the actual detector 
$N$ is finite, although very large, and the correction terms 
 might be tested experimentally. 
An attempt at the 
numerical estimate of the correction terms is presented.  
After the discussions in Sec.\ref{Discussions}, 
  general non-separable case is 
discussed in Appendix \ref{nonsepa}. 
In Appendixes \ref{Phibaaa},\ref{Phiabc},\ref{integX},
 some of calculational details are shown and in \ref{Neglect},
 the order estimation of neglected terms is given.  

\section{\label{stationarity}Stationary phase and macrovariable}

\subsection{\label{macro}Macroscopic system and macovariable}

Before constructing the model of measurement,
 the properties of macro and microvariables 
have to be elucidated, since the quantum mechanical detection
 process consists of an interplay between the two. 
Consider a macroscopic system, which contains a large number of microscopic
 particles with coordinates $x_i$~($i=1,2,\cdots,N$),
 all of which are assumed to have 
the same mass $m$. To make formulas simple, 
we work below in one dimension, extension to three dimensional case being 
straightforward.  
Now an extensive quantity grows up with the size 
of the system. 
As has been introduced in \cite{fukuda} in the case of field theory,
 a macrovarible of $N$ particle system is an intensive quantity 
defined by dividing an 
extensive quantity by $N$. 
The center of mass, $X=\sum_{i=1}^N x_{i}/N$ is the simplest example. 
  
To make above definition more precise in quantum case,
 let us recall that the quantum mechanical process is described by 
the path-integral form given by 
\be
\int \prod_{i=1}^N [dx_i ]{\rm e}^{\ri S/\hbar}
=\int\prod_{i=1}^N [dx_i ]{\rm e}^{\ri \int dt L/\hbar},
\label{SL}
\ee
where $S=\int dt L$ is the action functional of the system.  
The Lagrangian $L=L([x_i ],[\dot{x}_i ])$ is an extensive quantity,
 so it is $O(N)$. (($[x_i ]$ represents $x_i$'s collectively.) 
  For this statement to hold, the system has to be thermodydnamically normal,
  which holds when the interation 
among microdegrees is of short range
 and the particle density is finite over all space. 
(The statistical factor $1/N!$ has to be 
inserted properly 
for the system of identical 
particles.)

The macroscopic system behaves classically as a whole,
while the system at the same time contains atoms and molecules,
 which is described quantum mechanically.
 The stationary phase accompanying the macrovariable 
just realizes such a picture. The reason is simple; if we change one $x_i$ as 
$x_i \rightarrow x_i +a$ by a finite amount $a$, then 
 the change of the phase 
$S[x_i ]$ is of the order $a$. On the other hand, suppose 
the macrovariable $X$ changes by a finite amount $a$.  It means that 
macroscopic number $N$ of $x_i$'s are displaced by the order of $a$. 
Then $S[x_i ]$ in general shifts by $O(Na)$. 
 Now we integrate over all values $x_i$'s in (\ref{SL}),
 so when $N\rightarrow\infty$ 
only that point of $S[x_i]$ contributes which is stationary when $X$ varies. 
Since this holds for any time slice of the path-integral (\ref{SL}), 
one gets a smooth stationary trajectory of $X$.
 It describes the motion of a macroacopic body as a whole. 
On the other  hand, no stationary point exists for each $x_i$, so 
 every $x_i$ remains as a fluctuating quantum variable. 
Consider instead the limit $\hbar\rightarrow 0$. 
The change in $S[x_i]/\hbar$ under $x_i \rightarrow x_i +a$ is  
$O(a/\hbar)$ for each $i$, so every $x_i$ is determined by the 
 stationary equation, i.e. the Newtonian equation. 
 It is not the 
``classical limit'' as seen in the ordinary life.  

\subsection{\label{wavepart}Particle picture by the stationary phase}

The above statement is seen clearly if we 
take the separable case of the center of mass for $X$.
 (See  
Appendix \ref{nonsepa} for non-separable case.) 
 We write $x_i =X+x'_i$,
 where $x'_i$ is the coordinate measured from the 
center of mass and satisfies $\sum_i x'_i =0$.  
Now (\ref{SL}) becomes 
\be
\int[dX ]{\rm e}^{\ri S_1 [X]/\hbar}
\int\prod_{i=1}^N [dx'_i ]\delta(\sum_i x'_i )
{\rm e}^{\ri S_2 [x'_i ]/\hbar}.
\label{SL'}
\ee
The second factor describes the microscopic quantum phenomena and can be neglected, or integrated out,
 in the measurement thory, since we are interested only in $X$. 
The first factor accompanies the stationay phase since, as discussed above, 
 $S_1 [X]$ is proportional to $N$; $S_1 [X] =Ns_1 [X]$.  For large $N$,  
 the $X$ integration is dominated by the solution of functional stationary 
equation  
 $\delta s_1 [X]/\delta X(t)=0$.  
The fluctuation of $X$ is suppressed by the phase cancellation and a single 
smooth trajectory $X^{\rm st}(t)$ is  
selected by the constructive phase coherence among paths near $X^{\rm st}(t)$.  
Although the fluctuation of $X$ defines the wave function $\Psi(X,t)$ 
for finite $N$, once 
 $N$ becomes infinity,
$X$ reduces to the variable of a point-like particle. 
These statements are based on the following well-known formula.
 Let $f(X)$ be a function of $X$ 
having the stationary point at $X_0$, then we have
\be
\lim_{N\rightarrow \infty}{\rm e}^{\ri N f(X)}={\rm e}^{\ri N f(X_0 )}
\sqrt{\frac{2\pi\ri}{Nf''(X_0 )}}\delta(X-X_0 ).
\label{formula}
\ee
Consider here the Feynman  kernel $K(X,T;Y ,0)$, which 
connects the wave functions at different times;  
\be
\Psi(X,T)=\int K(X,T;Y ,0)\Psi(Y ,0)dY.  
\label{PsiK}
\ee
Applying (\ref{formula}) at every time slice from $t=0$ to $t=T$, the kernel is seen to  
 contain a factor 
$\delta(X-X^{\rm st}(Y,T))$, where $X^{\rm st}(Y,T)$ passes $Y$ at $T=0$. 
(Initial velocity depends on the form of $\Psi(X,0)$.) 
Thus, each point $X$ on  
 the wave function $\Psi(X,0)$ is just transported 
 along $X^{\rm st}(Y,T)$,
 as opposed to the Huygens picture of wave mechanics. 
 Our $\Psi(X,T)$ here represents the mixed state. 
If we choose $|\Psi(Y,0)|^2 =\delta(Y-X_I )$, then 
 $|\Psi(X,T)|^2 =\delta(X-X^{\rm st}(X_1 ,T))$;  
the wave function has a non-diffusive sharp peak, representing 
 the density of a point-like particle.
 Explicit examples appear later. 

To discuss other local densities, 
 let us discretize the time with the interval $\Delta t$. 
Then the fluctuating momentum operator \cite{Feynman}
 $P=(\hbar/\ri) \partial /\partial X(t)=Nm (X(t+\Delta t)-X(t))/\Delta t$ becomes 
the classical expression $Nm\dot{X}(t)$ evaluated along $X^{\rm st}(t)$ 
as $N\rightarrow\infty$.
 The quantum mechanical expression of the momentum or the energy e.t.c.  
 reduces to the corresponding classical density; 
\bea*
&&
\Psi^{*}(X,t)(~1,~P, ~P^2 /2Nm~)\Psi(X,t)~~\longrightarrow~~
\\&&
~~~~~(~1,~Nm\dot{X}(t),~Nm\dot{X}^2 /2~)\delta(X-X^{\rm st}(t)).
\eea* 

\subsection{\label{measure}Measurement by macrovariable}

Suppose we measure the object operator $O$ by the interaction Hamiltonian 
 $H_I (O,[x_i ],[p_i ])$.  Then the total Hamiltonian is the sum of three terms;
\be
H=H_O +H_D ([x_i ],[p_i ])+H_I (O,[x_i ],[p_i ] ), 
\label{HODI}
\ee
where $H_O$ is the object Hamiltonian, $H_D$
  that of the detector.  As stated in the Introduction,  
$O$ interacts with many particles in the detector 
 and $X$ is so chosen that it includes (almost) all of them;
\bbb
&&\hspace{-0.4cm}
 H_I =\sum_{i=1}^N h(O,x_i ,p_i ),~~~~X=\sum_{i=1}^N g(x_i ,p_i  ) /N
\label{general}
\eee
 with some functions $h$ and $g$. 
Here and hereafter, $N$ is the number of particles
 involved in (\ref{general}). 
 $H_D$ in (\ref{HODI}) is the Hamiltonian of these particles 
and is $O(N)$ together with $H_I$. Arguments of Secs \ref{macro}, \ref{wavepart} can be applied to $X$ thus defined. 
 
One comment here. If one can find any parameter $\alpha$ which 
produces the stationary phase for some variable $\eta$, then $\alpha$ and $\eta$ 
 can be used in place of $N$ and $X$. 
 Then, the macrocity will not be required for the measurement. 

\subsection{\label{Freeze}Freezing the object state}

Now in general, $O$ does not commute with $H_O$,
\begin{equation}
[H_O ,O]\neq 0.
\label{commute}
\end{equation}
So, the eigenstates of $O$ always fluctuates among different states.
 To measure such a fluctuating object,   
we have to suppress the size of the fluctuation somehow. 
This is done by taking $H_I =O(N)$; since $H_O =O(1)$, we can neglect $H_O$ for $N\rightarrow\infty$ 
 and adopt $H=H_D +H_I $
 as the Hamiltonian. Thus, 
taking the representation which
 diagonalizes $O$, $O|a>=\lambda_a |a>$,   
 each eigen-state $|a>$ develops by $H_D +H_I (\lambda_a ,[x_i ],[p_i ] )\equiv H_D +H_{I,a}$ and 
remains in the same state $|a>$ as long as $H_I \neq 0$,  
 i.e. the detector is on. Precisely during this
 period of $H_I \neq 0$, the macrovariable can change its value since it evolves 
by $H_D +H_{I,a}$,
 which produces different stationary phases for differnt
 $\lambda_a$'s since $H_I$ is $O(N)$. Thus we get different values 
of $X$ depending on the microscopic state of the object.
 It is realized independently of the detailed forms of the functions 
$h$ or $g$ above, as long as both include large number of 
particles which interact with $O$. 
This is the amplification mechanism of 
the detection process in terms of the stationary phase.
Higher order correction terms for large $N$
 are given systematically through the expantion
 in powers of the off-diagonal 
elements of the object Hamiltonian $H_O$. Below, this is done explicitly 
based on the model Hamiltonian.

\section{\label{modell}The model}

\subsection{\label{hamilton}The Hamiltonian}

When we construct the model, above functions $h(O,x_i ,p_i  )$
 and $g(x_i ,p_i )$ has to be 
fixed. The simplest case is $h=-fOx_i$ and $g=x_i$
 ($f$ is the coupling strength.); then
\be
H_I =-\sum_{i=1}^{N}fOx_i =-fNOX.
\label{HIO}
\ee
Here $X$ is the center of mass of particles that interact with $O$.   
In the realistic detector, the photomultiplier for example, 
 complicated processes may happen. 
 An object 
  interacts with an atom in the 
detector (via exchanging a photon),
 ionizing an electron. It is accelerated by the electric field applied in certain direction, 
which interacts with another atom, ionizing a 
second electron and so force, until we have a macroscopic number of 
electorns, giving a signal as the current. Or a high energy object interacts with many atoms in the direction on its momentum, 
along which the track of ionized electrons is detected. 
 Above $H_I$ 
 simulates these processes
 by a direct interaction of $O$ and $N$ particles, which 
are in the direction of applied electric
 field or in the direction of the object momentum. So    
the problem can be simulated by one dimension, with 
 $X$ taken to be the center of mass of $x_i$ in that direction. 
Independently of the detailed form of the interaction, 
the essential point is that, although each electron
 receives a microscopic amount of energy, 
 the sum of them is $O(N)$, which affects the stationary
 phase of $X$. 
Since $H_D$ is $P^2 /2Nm=Nm\dot{X}^2 /2$ ($P=Nm\dot{X}$ is the total momentum) plus terms 
independent of $X$, 
the total Hamiltonian, the object plus detector, of our model is defined by 
\begin{eqnarray}
H&=&
H_O +Nm\dot{X}^2 /2 -NfOX\equiv H_O +H_N,
\label{HOHN}
\\&&
H_N =Nm\dot{X}^2 /2-NfOX.
\label{model}
\end{eqnarray}
We do not write $x'_{i}$ part hereafter, since the dependence on $X$ and $x'_{i}$ 
is factorized as (\ref{SL'}).

\subsection{\label{initial}The initial wave function}

 Below the eigen-values $\lambda_a$ of $O$ are assumed to be discrete and 
non-degenerate; 
$\lambda_a \neq \lambda_b$ if $a\neq b$.  Writing the  
eigenstate of $X$ as $|X>$ with continuous eigen-value $X$, the complete set of 
states of our model Hamiltonian is given by $|a>|X>\equiv |a,X>$. 
We also use the complete set spanned by $|x,X>$. 
 The eigen-function is then $\phi_a (x)=<x|a>$. 

Let the detector be switched on at $t=0$, 
and the state vector at $t=0$ is written as $|\Psi>=|\phi>|\Phi>$, which is
 the product of the object $|\phi>$ and of the detector $|\Phi>$.
 Expanding as $|\phi>=\sum_a C_a |a>$ into complete sets, 
the initial wave function is given by  
\begin{eqnarray}
&&
\Psi(x,X,t=0)=<x,X|\Psi>=<x|\phi><X|\Phi>,
\nonumber
\\&&
<x|\phi>=\sum_a C_a <x|a>=\sum_a C_a \phi_a (x),
\label{psia}
\end{eqnarray} 
The initial wave function of the detector $\Psi(X)$ is assumed to have 
a peak at some value of $X$, 
with the precision $\Delta$. 
To be explicit,  
we adopt a Gaussian type;
\begin{equation}
<X|\Phi>=(\pi\Delta^2 )^{-1/4}
\exp(-X^2 /2\Delta^2 ).
\label{gauss}
\end{equation}
$|\Psi(X)|^2$ becomes $\delta(X)$ when $\Delta\rightarrow 0$.  
The center position is at $X=0$ and the initial velocity 
$\dot{X}=(\hbar/Nm\ri )\partial/\partial X$ is also zero
 when $N\rightarrow \infty$. (Non-zero velocity $v$ 
 is obtained by multiplying 
$\exp(\ri NmvX/\hbar)$ to (\ref{gauss}).) 
 
When $N\rightarrow\infty$,
 the center of the peak traces a classical trajectory determined
 by $H_N$ keeping the width $\Delta$ constant. 
 Here $<\!\!X|\Psi\!\!>|^2$ represents the density matrix of a mixed state, 
with $\Delta$ repesenting the classical uncertainty.  
 If $N$ is large but finite, $X$ fluctuates and 
 the diffusion process comes in. 
The numerical consideration 
is given at the end of Sec.\ref{psi0},
 where we will see that, up to the order we are considering,
 the influence of the diffusion is negligible
 in the detection process
 and the dominant effect 
 comes from 
 the fluctuation of the object while 
the detector is on. 

Now the macroscopic limit is $N\rightarrow\infty$,
 with other quantities kept fixed. 
But in order to avoid the classical uncertainty 
and keep various formulas simple, we take the limit 
 $\Delta\rightarrow 0$
 of the coefficients of limitting expression obtained by $N\rightarrow\infty$. 
Such a limit, first $N\rightarrow\infty$, then $\Delta\rightarrow 0$,
 is denoted as $\Rightarrow$. 

\subsection{\label{expoffdiag}Time evolution and the expansion by 
the off diagonal elements of $H_O$}

The total wave function develops in time as 
\begin{eqnarray}
&&
\Psi(x,X,T)=<x,X|\exp(-\ri HT/\hbar)|\Psi>
\nonumber
\\&&
=<x,X|\exp(-\ri HT/\hbar)\sum_a C_a |a>|\Phi>.
\label{timeevo}
\end{eqnarray} 
When we expand in powers of $H_O$, we first sum up the 
diagonal term $(H_O )_{aa}$ exactly in every order of expansion. 
Then the expansion becomes the one in terms of the 
power of off-diagonal elements $(H_O )_{ab},~a\neq b$. To achieve this,  
we use the well-known formula; 
\bbb
&&
U(T)=\exp(-\ri HT/\hbar)=U_N (T)U_O (T),
\nonumber
\\&&
U_N (T)=\exp(-\ri H_N T/\hbar),
\nonumber
\\&&
U_O (T)={\rm T}\exp\bigg(-\ri\int_{0}^T ds H_O (s)/\hbar\bigg),
\label{U_O}
\\&&
H_O (s)=U_N^{\dagger}(s)H_O U_N (s).
\eee
In (\ref{U_O}), ${\rm T}$ implies the time ordering operation.
To get the desired expansion, 
consider
\bbb
&&
H_O (s)|a>=U^{\dagger}_N (s)H_O U_N (s)|a>
\nonumber
\\&&
=U^{\dagger}_{N,a}(s) H_O |a> U_{N,a}(s)
\nonumber
\\&&
=U^{\dagger}_{N,a}(s) 
(H_O )_{aa}|a> U_{N,a}(s)
~+~{\rm off~\!diagonal~\!part}
\nonumber
\\&&
=(H_O )_{aa}|a>+{\rm off~diagonal~part}.
\label{diag}
\eee
Here we have introduced
\bbb
&& U_{N,a}(s)=\exp(-\ri H_{N,a}s/\hbar),
\label{Uas}\\&&
H_{N,a}=
 Nm\dot{X}^2 /2 -Nf_a X,~~~~~f_a =f\lambda_a .
\label{fa}
\eee
Thus, when we sum up the diagonal elements of $H_O$, $H_O$ can be treated as
 a c-number and the diagonal parts are summed up into the phase. 
Thus we can use the formula
\bbb
&&
U_N (T)U_O (T)|a>=U_{N,a}(T)\exp(\ri \theta_a T)|a>,
\nonumber
\\&&
\theta_a =-(H_O )_{aa}/\hbar.
\label{theta}
\eee
Now we concentrate on the defining equation of the T-product (\ref{U_O}).
 It is an infinite 
product of the term 
$\exp (-\ri H_O (s)\Delta s/\hbar)$ in the infinitesimal time interval 
$\Delta s$. 
When it is evaluated by off-diagonal elements $[H_O ]^{\rm nd}$,
 we write $\exp (-\ri H_O (s)\Delta s/\hbar
 \sim 1+(-\ri H_O (s)\Delta s/\hbar$. 
Using eqs.(\ref{diag}) and (\ref{theta}), 
 the expansion thus obtained becomes   
\be
<b,X|U_N (T)U_O (T)|a,Y>\equiv \sum_{k=0}^{\infty}U^{(k)}_{ba}(X,Y),
\label{exp}
\ee
where $U^{(k)}$ involves $k$-th power of $[H_O ]^{\rm nd}$.
 The results of first few terms are shown below; ( 
 $[U_O (T)]^{\rm d}$ contains the diagonal parts only, while 
$[H_O ]^{\rm nd}$ off-diagonal parts. )
\bbb
&&
U^{(0)}_{ba}(X,Y)=<b,X|U_N (T)|a,Y>
\nonumber
\\&&
~~~~~=\delta_{ab}\exp(\ri\theta_a T)<X|U_{N,a} (T)|Y>,
\label{U0}
\\&&
U^{(1)}_{ba}(X,Y)
=
<b,X|U_N (T)(-\ri/\hbar)
\nonumber
\\&&
~~~~~\times\int_0^T ds[U_O (T-s)]^{\rm d}[H_O (s)]^{\rm nd}
[U_O (s)]^{\rm d}|a,Y>
\nonumber
\\&&=
(-\ri/\hbar)(H_O )_{ba}\int_0^T ds \exp\{\ri(\theta_b (T-s)+\theta_a s)\}
\nonumber
\\&&~~~~~\times<X|U_{N,b}(T-s)U_{N,a}(s)|Y>. 
\label{U1}
\eee
 In (\ref{U1}) and in what follows, 
 $([H_O ]^{\rm nd})_{ba}$ is written simply as
 $(H_O )_{ba}$ for the notational 
simplicity so $a\neq b$ is implied.
The factor $\delta_{ab}$ in (\ref{U0}) 
 implies that the object does not fluctuate in the lowest order
 while the detector is on.   
By a similar manipulation, 
\bbb
&&
U^{(2)}_{ba}(X,Y)
=
(-\ri/\hbar)^2 \sum_c (H_O )_{bc}(H_O )_{ca}
\nonumber
\\&&
\times\int_0^T ds' \int_0^{s'} ds
\exp\{\ri(\theta_b (T-s')+\theta_c (s'-s))+\theta_a s\}
\nonumber
\\&&
\times
<X|U_{N,b}(T-s')U_{N,c}(s'-s)U_{N,a}(s)|Y>.
\label{U2}
\eee
The wave function has the corresponding expansion
\bea*
&&
\Psi(x,X,T)
=<x,X|U_N (T)U_O (T)\sum_{a} C_a |a,\Phi>
\\&&=\int dY\sum_{b,a} <x|b>C_a \sum_{k=0}^{\infty}U^{(k)}_{ba}(X,Y)<Y|\Phi>
\\&&
\equiv\sum_b <x|b>\sum_{k=0}^{\infty}\Psi^{(k)}(b,X,T)
=\sum_{k=0}^{\infty} \Psi^{(k)}(x,X,T).
\label{Psik}
\eea*
The wave function in the $|a>$ representation  
\be
 \Psi(b,X,T)=<b,X|U(T)|\Psi>
=\sum_{k=0}^{\infty}\Psi^{(k)}(b,X,T)
\label{wavefun}
\ee
has been introduced and each $\Psi^{(k)}$ has $k$-th power of $[H_O ]^{\rm nd}$. 
In case $[H_O ,O]=0$, ~$(H_O )_{ba}(\lambda_a -\lambda_b )=0$ follows, so one gets 
$[H_O]^{\rm nd}=0$. Thus,  
in (\ref{exp}) or (\ref{wavefun}) only the lowest term with $k=0$ is non-vanishing. 

\subsection{\label{signal}The signal function}

In the real experiment, the object 
 is not actually observed, 
so let us define the signal function by integrating (summipng up) 
 $|\Psi|^2$
 by $x$ ($b$);  
\bbb
&&
\hspace{-0.8cm}J(X,T)=\int dx |\Psi(x,X,T)|^2 =\sum_b |\Psi(b,X,T)|^2 
\label{J}
\\&&
~~~~=\sum_{k=0}^{\infty}J^{(k)}(X,T),
\\&&
\hspace{-0.9cm}J^{(0)}(X,T)=\sum_b |\Psi^{(0)}(b,X,T)|^2 ,
\label{zeroth}
\\&&
\hspace{-0.9cm}J^{(1)}(X,T)=\sum_b \Psi^{(0)*}(b,X,T)\Psi^{(1)}(b,X,T)+{\rm c.c.},
\label{1st}\\&&
\hspace{-0.9cm}J^{(2)}(X,T)=J^{(2)}_1 (X,T)+J^{(2)}_2 (X,T),
\nonumber\\&&
\hspace{-0.9cm}J^{(2)}_1 (X,T)=\sum_b \Psi^{(1)*}(b,X,T)\Psi^{(1)}(b,X,T), 
\label{2nd1}\\&&
\hspace{-0.9cm}J^{(2)}_2 (X,T)=\sum_b 
\Psi^{(0)*}(b,X,T)\Psi^{(2)}(b,X,T)+{\rm c.c.}.
\label{2nd2}
\eee
Here $J^{(k)}(X,T)$ is of the order of $([H_O ]^{\rm nd})^k$. 
Let us calculate $J^{(0,1,2)}(X,T)$ and $\Psi^{(0,1,2)}$ successively.

\subsection{\label{psi0}$\Psi^{(0)}(b,X,T)$}

The lowest term is calculated by (\ref{theta}), (\ref{U0}) and (\ref{wavefun}) as 
\bea*
&&
\Psi^{(0)}(b,X,T)=\sum_a C_a <b,X|U_{N} (T)|a,\Phi>
\\&&
=\int dY  
\exp(\ri\theta_b T) C_b <X|U_{N,b} (T)|Y><Y|\Phi>. 
\eea*
Now we insert (\ref{gauss}), 
and use the following result of the Feynman kernel \cite{Feynman} for 
the Hamiltonian $H_{N,a}$ of (\ref{fa}); 
\bbb
&&
<X|U_{N,a} (T)|Y>=
\sqrt{\frac{Nm}{2\pi\ri \hbar T}}
{\rm e}^{\ri N\Theta_a /\hbar}{\rm e}^{\ri \tilde{\theta}_a },
\label{UNa}
\\&&
\Theta_a =\frac{m(X-Y-\xi_a (T))^2}{2T},
\nonumber
\\&&
\xi_a (T)=f_a T^2 /2m,
\label{xia}
\eee
$\xi_a (T)$ is the classical change of $X$ during $T$ in the presence of 
the constant force $f_a$, with the initial condition $X=\dot{X}=0$.
 $\tilde{\theta}_a$ is the
 classical action for this motion. 
By (\ref{formula}), 
\bbb
&&
 \hspace{-0.6cm}
\lim_{N\rightarrow\infty}<X|U_{N,a} (T)|Y>=
\delta (X-Y-\xi_a (T))
{\rm e}^{\ri \tilde{\theta}_a }. 
\label{UNaa}
\eee
Apart from the phase, the whole 
wave function develops as a parallel transport;  
\bbb
&&
\Psi^{(0)}(x,X,T)={\rm e}^{\ri \tilde{\theta}_a }\Psi^{(0)}(X-\xi_a (T),0).
\label{PP}
\eee
For the discussions below, the $Y$-integration is done 
for general $N$. After a simple Gaussian integral, we get
\bbb
&&
\Psi^{(0)}(x,X,T)=
C_b {\rm e}^{\ri\theta_b T}\left(
\frac{1}{\pi\Delta^2}\right)^{1/4}\sqrt{\frac{Nm}{\ri\hbar TD}}{\rm e}^{\ri \tilde{\theta}_b}
\nonumber\\&&
\hspace{-0.4cm}\times\exp\left[
-\frac{N^2 m^2}{T^2 \hbar^2}
\frac{(X-\xi_b (T))^2}{2D}
+\frac{\ri Nm(X-\xi_b (T))^2}{2\hbar T}
\right],
\nonumber
\\&&
D=\frac{1}{\Delta^2}-\frac{\ri Nm}{\hbar T}.
\label{psi^0}
\eee
Taking the absolute square,  
\bbb
&&\hspace{-1.0cm}
|\Psi^{(0)}_b (X,T))|^2 
=
 |C_b |^2 \sqrt{\frac{1}{\pi\Delta^2 \rho}}\exp\left[
-\frac{(X-\xi_b (T))^2}{\Delta^2 \rho}
\right].
\label{rhoo}
\eee
Here we have written
\be
DD^* =\frac{N^2 m^2}{\hbar^2 T^2}\rho,~~~~~~
\rho\equiv 1+\frac{T^2 \hbar^2}{\Delta^4 N^2 m^2}.
\label{DD*}
\ee
The peak of $|\Psi^{(0)}(b,X,T)|^2$ traces the classical trajectory
 $X=\xi_b (T)$ and for large $N$,  
 $\rho=1+O(1/N^2 )$, 
so the effect of the broadening of the width due to the fluctuation of $X$,
 is $O(1/N^2)$ so 
setting  $\rho=1$ is may be allowed. See blow for the numerical study. 

Now, in order to map microscopically different channels $|a>\neq |b>$
 into a macroscopically distinguishable state, $\Delta$ has to be 
suffiently small compared with the distance of the different peaks;
$|~\!\xi_a (T)-\xi_b (T)|>\!\!>\Delta$. 
This is the requirement for the detector, which is assumed 
to be the case.
 To get the numerical value  
 of $T$ required for producing a signal, we estimate 
\bea*
&&
|~\!\xi_a (T)-\xi_b (T)|\sim \frac{f_a T^2}{2m}>\!\!>\Delta. 
\eea*
Thus we get $T\!>\!\!>\sqrt{2m\Delta/f_a}$.
Let $a$ be the atomic scale length,
 then $f_a a=f\lambda_a a\sim fOa$ 
is of the atomic energy size (inonization energy, for instance)
 due to the interaction between the object and one
 particle in the detector. 
If we set
 rather arbitrarily $f_a a$ =1~\!eV=$1.6\times 10^{-12}$erg, 
 $\Delta =10^{-3}$~\!cm and 
 take $a=10^{-7}$~\!cm, then for the case of the electron 
($m=$9.1$\times 10^{-28}$g),
\bbb
&&
T\!>\!\!>\sqrt{\frac{2m\Delta a}{f_a a}}\sim 10^{-12}~\!{\rm sec.}
\label{10-12}
\eee
This is quite a small number, which  
 does not change 
much for the proton 
($m=$1.7$\times 10^{-24}$g) 
and for somewhat larger or samller $\Delta$.  

Next, we try to estimate the magnitude of $N$, for which the diffusion process 
during the measuement time $T$ can be neglected. 
 By (\ref{rhoo}), 
$\rho-1<\!\!<1$,
 or equivalently  $N\!>\!\!>\hbar T/(\Delta^2 m)$  
has to be satisfied. 
This follows also from the uncertainty relation. 
If we adopt $\Delta =(10^{-3},10^{-4})$cm, 
then we get $N\!>\!\!>(10^6 ,10^8 )T$
 for the electron, 
 $N\!>\!\!>(10^3 ,10^5 )T$ for the proton. 
 ($T$ is measured in sec.)
 Since $T$ given in 
 (\ref{10-12}) is quite samll, we conclude  
that the diffusion effect in the 
detection process is totally negligible. We set $\rho=1$ hereafter. 
When we include higher orders of $[H_O ]^{\rm nd}$, 
non-trivial constraint on $N$ will emerge, see Sec.\ref{numerical}. 

\subsection{\label{0thsignal}The signal function in zeroth order}

$J^{(0)}(X,T)$ is given in (\ref{zeroth}). In the limit $\Rightarrow $, we get  
 the ideal situation in the  measurement. Denoting by $\longrightarrow$ the 
time evolution after the detector is switched on, we get
\bbb
&&
 J^{(0)}(X,0)=\sum_b |C_b |^2 \delta(X)=\delta(X)
\label{deltaX}
\\&&~\longrightarrow~J^{(0)}(X,T)=\sum_b |C_b |^2  
\delta(X-\xi_b (T))
\label{ideal}
\eee 
Note that the signal function becomes the classical
 density of a point particle moving along 
 $\xi_b (T)$.  
 By ideal, we mean that the above result
 is in conformity with the usual 
the quantum mechanical rule; integrating by $X$ in the samll region 
$R_b =(\xi_b (T)-\delta,~\! \xi_b (T)+\delta)$, ($\delta>0$), 
\be
\int_{R_b} J^{(0)}(X,t)dX=|C_b |^2 .
\label{intR}
\ee
We say that it is the probability for $X$ to stay in $R_b$, which in turn implies 
that the probability of the object to be in the state $|b>$ is $|C_b |^2$, 
since the mapping 
$|b>\leftrightarrow \xi_b (T)$ is one to one by the stationary phase mechanism. 

Written by the wave function symbolically, the ideal 
measuring process is expressed as
\bea*
&&
\Psi(x,X,0)=\sum_a C_a <x|a>\sqrt{\delta(X)}~
\longrightarrow~
\\&&
\Psi(x,X,T)=\sum_a C_a <x|a>  
{\rm e}^{\ri (\theta_a T+\tilde{\theta}_a )}\sqrt{\delta(X-\xi_a (T))}.
\eea*
Thus the object stays in the same state.  
(Suare-root of the delta-function is ill-defined so 
we need some reguralization.)

\section{\label{higher} Higher orders}
\subsection{\label{psi1} $\Psi^{(1)}(b,X,T)$}

As is given in (\ref{U1}), 
$<X|U_{N,b} (T-s)U_{N,a} (s)|Y>$ has to be evaluated.  
This is the evolution kernel 
defined by the Hamiltonian $H(t)=Nm\dot{X}^2 /2-Nf(t)X$, where 
\be
f(t)=
\left\{
\begin{array}{l}
f_a ~;~~~~{\rm for}~~~0<t<s, \\
f_b ~;~~~~{\rm for}~~~s<t<T.
\end{array}
\right.
\label{f(t)}
\ee
Now we apply the formula for this process \cite{Feynman} 
\bea*
&&\hspace{-0.4cm}
<Y|U_{N,b} (T-s)U_{N,a} (s)|X>
=\sqrt{\frac{Nm}{2\pi\ri \hbar T}}\exp\left(\frac{\ri}{\hbar}NS_{ba}\right),
\\&&
S_{ba}=\frac{m(Y-X)^2 }{2T}
\\&&~~~~~~~
+\frac{Y}{T}\int_0^T dtf(t)t+
\frac{X}{T}\int_0^T dtf(t)(T-t)
\\&&~~~~~~~
-\frac{1}{Tm}\int_0^T dt\int_0^T dt' f(t)f(t')(T-t)t' .
\eea*
Using (\ref{f(t)}), we get after a straightforward calculation, 
\bbb
&&
S_{ba}=\frac{m}{2T}\bigg(Y-X-\xi_{ba}(T,s)\bigg)^2 
\\&&~~~~~~~~~~~~~~~+XQ_{ba}(T,s)+P_{ba}(T,s),
\nonumber\\
&&
Q_{ba}(T,s)=f_a s +f_b (T-s),
\label{Qba}\\&&
P_{ba}(T,s)=-\frac{f_a^2}{6m}s^2 (3T-2s)
\nonumber\\&&~~~~~~~~~~~~~-\frac{f_b^2}{6m}(T-s)^3 -\frac{f_a f_b}{2m}s(T-s)^2 ,
\label{Pba}
\\&&
\xi_{ba}(T,s)=\frac{f_a}{2m}(2Ts-s^2 )+\frac{f_b}{2m}((T-s)^2 .
\label{xiba}
\eee
Here, $\xi_{ba}(T,s)$ is the classica change of $X$ during $T$ 
under the force $f(t)$. Note that $\xi_{ba}(T,0)=\xi_b (T)$ and $\xi_{ba}(T,T)=\xi_a (T)$,
 in comformity with the fact that $s$ 
is the time $H_O$ acted, making the transition from $|a>$ to $|b>$. 
One can confirm again that 
$P_{ba}(T,s)+XQ_{ba}(T,s)$ coincides with 
the classical action integral along $\xi_{ba}(T,s)$. 
Using (\ref{gauss}), (\ref{wavefun}) and applying the operation 
$\int dY \sum_a C_a <Y|\Phi>$, we finally obtain
\bbb
&&
\Psi^{(1)} (b,X,T)
\\&&
=\left(\frac{1}{\pi\Delta^2}\right)^{1/4}\sqrt{\frac{Nm}{2\pi\ri\hbar T}}\sqrt{\frac{2\pi}{D}}
\sum_a (-\ri/\hbar)(H_O )_{ba}C_a 
\nonumber\\
&&
\times
\int_0^T ds\times\exp\ri\{\theta_a s+\theta_b (T-s)\}
\nonumber\\&&
\times
\exp\left[
-\frac{N^2 m^2 R_{ba}(X,T,s)^2}{2\hbar^2 T^2 D}+\frac{Nm\ri R_{ba}(X,T,s)^2}{2\hbar T}
\right.
\nonumber\\&&
~~~~~~~~~~~~\left.+\frac{\ri}{\hbar}N\{XQ_{ba}(T,s)+P_{ba}(T,s)\}\right],
\label{psi^1}\\&&
R_{ba}(X,T,s)=X-\xi_{ba}(T,s),
\label{Rba}
\eee
Note that $R_{ba}(X,T,0)=X-\xi_{b}(T)$. 

\subsection{\label{J1}Signal function $J^{(1)}(X,T)$}

By (\ref{1st}), (\ref{psi^0}) and (\ref{psi^1}), one obtains 
\bbb
&&
J^{(1)}(X,T)
=\left(\frac{1}{\pi\Delta^2 \rho}\right)^{1/2}
(-\ri/\hbar)\sum_{ab}C^*_b (H_O )_{ba}C_a 
\nonumber\\&&
~~\times \int_0^T ds\times\exp\ri\{(\theta_a -\theta_b )s\}\exp{\Phi}_{ba}
~+~{\rm c.c.}. 
\nonumber
\eee
The explicit expression of $\Phi_{ba}$ is shown in 
 (\ref{Phiccc}), (\ref{omegaaa}) of Appendix \ref{Phibaaa};
The result is 
\bbb
&&
\Phi_{ba}=
-\frac{1}{\Delta^2}\bigg(X-\displaystyle 
\frac{\xi_{ba}(T,s) +\xi_{ba}(T,0)}{2}\bigg)^2 
\nonumber
\\&&\hspace{-0.5cm}
-\frac{1}{4\Delta^2 }(\xi_{ba}(T,s)-\xi_{ba}(T,0))^2 
+\frac{\ri N}{\hbar}\omega_{ba}(X,T,s).
\label{PhiPhi}
\eee
Here $\omega_{ba}(X,T,s)$ is given by 
\bea*
&&
\omega_{ba}(X,T,s)
=X(f_a -f_b )s 
-\frac{f_a^2}{6m}s^2 (3T-2s)
\\&&
~~-\frac{f_b^2}{6m}\{(T-s)^3 -T^3 \} -\frac{f_a f_b}{2m}s(T-s)^2 .
\eea*
When $N\rightarrow\infty$, the integration by $s$ is dominated by the 
stationary point, satisfying 
\bbb
&&
\hspace{-0.8cm}0=\frac{d\omega_{ba}(X,T,s)}{ds}
=(f_a -f_b )\bigg(X-\frac{f_b T^2 }{2m} 
\nonumber\\&&~~~~~~~~~~-\frac{(f_a -f_b )Ts}{m}
+\frac{(2f_a -f_b )s^2}{2m}\bigg).
\label{omegas}
\eee
Taking in advance the limit $\Delta\rightarrow 0$ into consideration, 
\bbb
&&
X-\displaystyle \frac{\xi_{ba}(T,s) +\xi_{ba}(T,0)}{2}=0,
\label{XXxi}
\\&&\hspace{-0.6cm}
\xi_{ba}(T,s)- \xi_{ba}(T,0)=(f_a -f_b )(2T-s)s=0
\label{Xxi}
\eee
have to be satisfied also. The only solution of (\ref{Xxi}) in 
the range $0\leq s\leq T$ is $s=0$, which also satisfies (\ref{omegas}). Then 
 (\ref{XXxi}) becomes
 $X-f_b T^2 /2m=0$. 
Thus, as a function of $X$, $\Psi^{(0)*}\Psi^{(1)}$ has a peak at $X=\xi_b (T)$. 
These facts are expected; suppose the peak of 
$\Psi^{(0)}(X,T)$ is at $\xi_b (T)$. In order to get non-zero $J^{(1)}(X,T)$, 
the peak of $\Psi^{(1)}(X,T)$ shoud also be at $\xi_{b}(T)$. 
This can be 
realized if and only if the transition caused by $[H_O ]^{\rm nd}$ 
from the state $|a>$ to $|b>$  
 occurrs at $s=0$. Then the $X$-integration in
  $\Psi^{(0)*}\Psi^{(1)}$ is dominated near $X=\xi_b (T)$.  
To perform the calculation, 
 consider the region near $s=0$, $X=\xi_b (T)$;
\bbb
&&
0=\frac{d\omega_{ba}(X,T,s)}{ds}
\nonumber\\&&\hspace{-0.6cm}
\sim (f_a -f_b )\left(X-\frac{f_b T^2}{2m} -\frac{f_a -f_b }{m}Ts\right)+O(s^2 ). 
\label{zero}
\eee
Thus, near $s=0$, the stationry tarjectory $s=s(X)$ and the second derivative 
(which is $X$-independent) becomes 
\bbb
&&
\hspace{-0.6cm}
s=s(X)=\frac{m(X-\xi_b (T))}{(f_a -f_b )T}+O((X-\xi_b (T))^2 ),
\label{sX}
\\&&
\frac{\partial^2 \omega_{ba}(X,T,s)}{\partial s^2}
=-\frac{(f_a -f_b )^2 T}{m}.
\label{partial}
\eee
Note that the second derivative is a constant. 
Keeping $X$ fixed, $s$-integrarion is first performed by expanding 
$\omega_{ba}(X,T,s)$ around $s=s(X)$
\bea*
&&
\omega_{ba}(X,T,s)=\omega_{ba}(X,T,s(X))
\\&&~~~~~~~~~~~~-(T/2m)(f_a -f_b )^2 (s-s(X))^2 
+\cdots\cdots.
\eea*
In this way, we get
\bbb
&&
\hspace{-0.7cm}\int_0^T ds\exp(\ri N \omega_{ba}(X,T,s)/\hbar)
\nonumber
=\exp(\ri N\omega_{ba}(X,T,s(X))/\hbar)
\\&&
~~~~~\times\sqrt{\frac{-2\ri\pi\hbar m}{N(f_a -f_b )^2 T}}\bigg(1+O(1/\sqrt{N})\bigg).
\label{sint}
\eee
The above expression is a function of $X$. Now we consider its asymptotic 
functional form when $N\rightarrow \infty$.
 The stationary point is given  by 
\bea*
&&
\hspace{-0.3cm}0=\frac{d\omega_{ba}(X,T,s(X))}{dX}=(f_a -f_b )s(X). 
\eea*
Therefore $X=\xi_{b}(T)$ and 
\bea*
&&
\omega_{ba}(X=\xi_b (T),T,s(\xi_b (T)))=\omega_{ba}(\xi_b (T),T,0)=0,
\\&&
\frac{d^2 \omega_{ba}(T,s(X))}{dX^2}=(f_a -f_b )\frac{ds(X)}{dX}
=\frac{m}{T}.
\eea*
Thus for large $N$, one can write
\bbb
&&
\exp(\ri N\omega_{ba}(X,T,s(X))/\hbar)
\nonumber\\&&
=
\exp\left(\frac{\ri N}{\hbar}\right)\left(\frac{m}{2T}(X-\xi_b (T) )^2 +
O((X-\xi_b (T) )^3 )\right),
\nonumber\\&&
~~\Longrightarrow~~\frac{1}{2}
\sqrt{\frac{2\ri \pi \hbar T}{Nm}}\bigg(
\delta(X-\xi_b (T))+O(1/\sqrt{N})\bigg).
\label{omegast}
\eee
The factor $1/2$ in front appears by the following reason. By (\ref{sX}) and by $s>0$, it follows that 
$X>\xi_b (T)$ ($X<\xi_b (T)$) if 
$f_a >f_b$ ($f_a <f_b$) along the stationary trajectory. Therefore, in either case, 
$X=\xi_b (T)$ is the end point of the $X$-integration and using the
 formula $\int_0^{\infty} dx\delta(x)=1/2$, eq.
(\ref{omegast}) follows. 

Other factors in (\ref{PhiPhi}) not multiplied by $N$ are unity for large $N$, when the stationary value is inserted.  
In fact, consider
\bbb
&&
\hspace{-0.6cm}\exp\left\{-\frac{1}{\Delta^2 }\left(
X-\frac{\xi_{ba}(T,s(X))+\xi_{ba}(T,0)}{2}\right)^2 \right\}
\label{1}
\\&&
\times\exp\left\{-\frac{1}{4\Delta^2 }\left(\xi_{ba}(T,s(X))-\xi_{ba}(T,0)\right)^2 \right\}. 
\label{2}
\eee
Since (\ref{omegast}) says that $X-\xi_b (T)=O(1/\sqrt{N})$ and
 $s(X)\sim X-\xi_b (T)$, one can estimate for large $N$ as  
\bea*
&&
\xi_{ba}(T,s(X))-\xi_{ba}(T,0)=O(s(X))
\\&&~~~~~~~~~~=O(X-\xi_b (T))=O(1/\sqrt{N}),
\\&&
\hspace{-0.3cm}
 X-\frac{\xi_{ba}(T,s(X))+\xi_{ba}(T,0)}{2}
=X-\xi_b (T)\\&&
~~~~~~~~~~-\xi'_{ba}(T,0)s(X)/2 -\xi''_{ba}(T,0)s(X)^2 /4+\cdots
\\&&~~~~=-\xi''_{ba}(T,0)s(X)^2 /4+\cdots.
\eea*
This is $O(X-\xi_b (T))^2 =O(1/N)$. 
Therefore, both factors of (\ref{1}) and (\ref{2}) becomes unity as $N\rightarrow \infty$. 

Collecting (\ref{sint}), (\ref{omegast}),
 and adding the term with complex conjugate, 
 we arrive at
\bbb
&&
J^{(1)}(X,T)=\sum_b \Psi^{(0)*}_b (X,T)\Psi^{(1)}_b (X,T)+{\rm c.c.}
\nonumber\\&&~~~~~~~~~~~~~~
=\frac{1}{N}\sum_b K^{(1)}_b \delta\bigg(X-\frac{f_b^2 T}{2m}\bigg),
\label{J^(11)}
\\&&\hspace{-0.6cm}
K^{(1)}_b 
=\frac{\sqrt{\pi}}{\Delta}
\sum_{a}\frac{2{\rm Im}~\!C^*_b (H_O )_{ba}C_a}{|f_a -f_b |}~+~O(1/N).
\label{J^(1)}
\eee
This is the first order correction in $[H_O ]^{\rm nd}$ to the ideal case (\ref{ideal}). 
In Appendix \ref{integX}, the result (\ref{J^(1)}) is checked by integrating over $X$ first 
and then by $s$. 
Note that $\sqrt{T}$ in (\ref{sint}) and (\ref{omegast}) are cancelled, 
so $K_b$ is independent of $T$ for each $b$. In this connection, see
Sec.\ref{Tindep}. 

\subsection{\label{normalization}Normalization}

The normalization $\int dXJ(X,T)=1$ leads to 
\bbb
&&
\int dXJ^{(0)}(X,T)=1, 
\nonumber\\&&
\int dXJ^{(k)}(X,T)=0~~(k=1,2,\cdots). 
\label{normal}
\eee
We can check $\int dXJ^{(1)}(X,T)=0$. Indedd, note that  
\bea*
&&
\int dXJ^{(1)}(X,T)=\frac{\sqrt{\pi}}{N\Delta}
\sum_{a,b}\frac{2{\rm Im}~\!C^*_b (H_O )_{ba}C_a}{|f_a -f_b |}. 
\eea*
Here, $1/|f_a -f_b |$ is real and symmetric 
 under $a\leftrightarrow b$. 
Then, we see that  
$\sum_{b\neq a}C^*_b (H_O )_{ba}C_a /|f_a -f_b |$ is a real number, so the imaginary part vanishes. 
   
\subsection{\label{1*1}Calculation of $J^{(2)}_1 (X,T)$}

Consider $J_1^{(2)}$ of (\ref{2nd1}).  It is expressed by
\bbb
&&
J^{(2)}_{1}(X,T)= \sum_b \Psi^{(1)*}_b (X,T)\Psi^{(1)}_b (X,T)
\nonumber\\&&\hspace{-0.3cm}
=\left(\frac{1}{\pi\Delta^2 \rho}\right)^{1/2}
\sum_{aa'b}C^*_{a'}(H_O)_{a'b}(H_O )_{ba}C_a 
\int_0^T ds'\int_0^T ds
\nonumber\\&&\hspace{-0.3cm}
\times
\exp\ri\{-(\theta_{a'} s'+\theta_b (s-s')-\theta_a s)\}
\exp~\!{\Phi}_{a'a;b}.
\label{Phia'ab}
\eee
${\Phi}_{a'a;b}$ is given in (\ref{Phia'a;b}) and (\ref{abc}) of Appendix \ref{Phiabc};
\bbb
&&\hspace{-0.7cm}
\Phi_{a'a;b}=
-\frac{1}{\Delta^2}
\bigg(X-\displaystyle \frac{\xi_{ab}(T,s) +\xi_{a'b}(T,s')}{2}\bigg)^2 
\nonumber
\\
&&\hspace{-0.7cm}
-\frac{1}{4\Delta^2 }(\xi_{ba}(T,s)-\xi_{ba'}(T,s'))^2 
+\frac{\ri N}{\hbar}\omega_{aa';b}(T,s),
\label{Phia'a;bb}
\\
&&
\omega_{a'a;b}(T,s)=\omega_{ba}(T,s)-\omega_{ba'}(T,s')
\label{abcc}
\\&&\hspace{-0.6cm}=
X(Q_{ba}(T,s)-Q_{ba'}(T,s'))+P_{ba}(T,s)-P_{ba'}(T,s').
\nonumber
\eee
The stationary equation in $s$ is identical to (\ref{omegas}); 
\bbb
&&
0=\frac{d\omega_{aa';b}(T,s,s')}{ds}=\frac{d\omega_{ba}(T,s)}{ds}
\eee
The solution is written as $s=s_{ba}(X)$.
Similarly, we have 
\bbb
&&
0=\frac{d\omega_{aa';b}(T,s,s')}{ds'}=-\frac{d\omega_{ba'}(T,s')}{ds'}
\nonumber\\&&
=-(f_{a'} -f_b )\bigg(
X-\frac{f_b T^2}{2m}-\frac{(f_{a'} -f_b )Ts'}{m}
\\&&~~~~~~~~~~~~~~~~-\frac{(2f_{a'} -f_b )s'^{2} }{2m}\bigg),
\label{fa'fb}
\eee
with the solution $s'=s'_{ba'}(X)$.
 Now $\Phi_{aa';b}(X,s,s')$ is expanded around $s=s_{ba}(X)$ and $s'=s'_{ba'}(X)$, 
and we calculate the second derivative at these points,  
\bbb
&&
M_{ba}\equiv \partial^2 \omega_{aa';b}/\partial s^2 
=
 -(f_a -f_b )^2 T/m 
\nonumber\\&&~~~-(f_a -f_b )(2f_a -f_b )s_{ba}(X) /m, 
\label{M}
\\&&
M_{ba'}\equiv\partial^2 \omega_{aa';b}/\partial s'^{2}
=(f_{a'} -f_b )^2 T/m 
\nonumber\\
&&
~~~+(f_{a'} -f_b )(2f_{a'} -f_b )s'_{ba'}(X) /m.
\label{M'}
\eee 
The reult of 
$s,s'$ integration is 
\bbb
&&
\int_0^T ds\int_0^T ds'
\exp\ri\{-(\theta_{a'} s'+\theta_b (s-s')-\theta_a s)\}
\nonumber\\&&
~~~~~~~~~~~~~\times\exp~\!\Phi_{aa';b}(X,s,s')
\nonumber\\&&\hspace{-0.7cm}
=\sqrt{\frac{(2\pi)^2 \hbar^2 }{N^2 M_{ba}M_{ba'}}}.
\exp~\!\Phi_{aa';b}(X,s_{ba}(X),s'_{ba'}(X)).
\label{MM}
\eee
Next task is to study the $X$-integration. For that purpose, it is convenient 
to use the following form for the factor appearing in (\ref{Phia'a;bb});
\bbb
P&\equiv&
 \exp\left[
-\frac{1}{\Delta^2}\bigg(X-\displaystyle 
\frac{\xi_{ba}(T,s) +\xi_{ba'}(T,s')}{2}\bigg)^2 \right]
\nonumber\\&&~~
\times\exp\left[-\frac{1}{4\Delta^2}(\xi_{ba}(T,s)-\xi_{ba'}(T,s'))^2 \right]
\nonumber\\
&=&
\exp\left[-\frac{1}{2\Delta^2}\bigg(X- \xi_{ba}(T,s)\bigg)^2 \right]
\nonumber\\&&~~
\times\exp\left[-\frac{1}{2\Delta^2}\bigg(X- \xi_{ba'}(T,s')\bigg)^2 \right].
\label{DeltaDekta}
\eee
Due to the structure of $\Phi_{aa';b}(X,s_{ba}(X),s'_{ba'}(X))$, 
the resulting dependence on $X$ differs for $a'=a$ and $a'\neq a$. 

\v
\n
\u{The case $a=a'$}

\v
\n
$\omega_{aa;b}(X,s_{ba}(X),s'_{ba'}(X))=0$ holds since for $a=a'$ 
$s_{ba}(X)=s'_{ba'}(X)\equiv s(X)$.  
Thus the result of the 
$X$-integration is a constant independent of $N$.
Consider the factor (\ref{DeltaDekta}) contained
 in $\Phi_{aa';b}$ of (\ref{Phia'a;bb}). 
Inserting the stationary value $s(X)=s_{ba}(X)$  in the first factor
 of the right-hand side of (\ref{DeltaDekta}), 
 we concentrate on $X-\xi_{ba}(s(X))$. 
The factor in the second 
parenthesis in (\ref{omegas}) 
 is rearranged as  
\bbb
&&
X-\frac{1}{2m}\bigg(f_b T^2  +2(f_a -f_b )Ts(X)
-(2f_a -f_b )s(X)^2 \bigg)
\nonumber
\\&&
=X-\xi_{ba}(T,s(X))+\frac{1}{2m}f_a s(X)^2 .
\eee
In this way, we get  
\be
X-\xi_{ba}(T,s(X))
=-(f_a /2m) s(X)^2 .
\label{s(X)zero}
\ee
Since we are considering $\Delta \rightarrow 0$ (after $N\rightarrow\infty$), 
$X-\xi_{ba}(T,s(X))\rightarrow 0$, implying $s(X)\rightarrow 0$. 
On the other hand, by (\ref{xiba}), one can approximate
\bea*
&&X-\xi_{ba}(T,s(X))
\\&&
~~= X-\frac{f_b T^2}{2m}
-\frac{(f_a -f_b )T}{m}s(X)
= 0.
\eea*
Soving this relation for $s(X)$ and 
inserting it back into (\ref{s(X)zero}), we conclude 
\bea*
&&\hspace{-0.3cm}
X-\xi_{ba}(T,s(X))=-\frac{f_a}{2m}\frac{m^2}{(f_a -f_b )^2 T^2} 
\left(X- \frac{f_b T^2}{2m}\right)^2 .
\eea*
Thus we obtain
\bea*
&&P
=\exp\left[-\frac{C_{ba}}{\Delta^2 T^4}
\left(X- \frac{f_b T^2}{2m}\right)^4 \right),~~
\\&&
~~~~~~~~
C_{ba}=\frac{(f^2_a}{4m^2}\frac{m^4}{(f_a -f_b )^4}.
\eea*
Here the following formula is adopted. With $C>0$,
\bea*
&&
\lim_{\Delta\rightarrow 0}
\exp\left[
-C\frac{(X-a)^4}{\Delta^2}\right]
=\frac{\gamma\sqrt{\Delta}}{C^{1/4}}\delta(X-a),
\eea*
where $\gamma\equiv \int_{-\infty}^{\infty} dz\exp(-z^4)$.
In this way, we arrive at 
\bea*
&&
P=\frac{\gamma\sqrt{\Delta}}{C^{1/4}}\frac{1}{2}
\delta\left(X-\frac{f_b T^2}{2m}\right).
\eea*
The factor $1/2$ is present for the same reason as given concerning
 (\ref{omegast}). Collecting all factors, the result
 for $J^{(2)}_{1;a=a'}$ is obtained 
as follows. In doing so, $s_{ba}(X)$ and $s_{ba'}(X)$ appearing in 
$M_{ba},~M_{ba'}$ of (\ref{M}),~(\ref{M'}) can be set to zero,
 which is inserted into (\ref{MM}). We use (\ref{DD*}) and set $\rho=1$. 
\bbb
&&
J^{(2)}_{1;a=a'} 
=\sum_{a=a'b}\delta_{aa'}(-\ri/\hbar)(\ri/\hbar)C^*_{a'}(H_O )_{a'b}
(H_O )_{ba}C_a 
\nonumber
\\
&&
~~~~\times\sqrt{\frac{1}{\pi\Delta^2}}\frac{Nm}{2\pi\hbar T}\frac{2\pi}{\sqrt{DD^*}}
\nonumber\\&&
~~~~\times\frac{1}{N}\frac{\pi\hbar 2m}{(f_a -f_b )^2 T}
\frac{\gamma\sqrt{\Delta}T}{C^{1/4}_{ab}}\frac{1}{2}
\delta\bigg(X-\frac{f_b T^2}{2m}\bigg)
\nonumber\\&&
~~~~~~~~~=\frac{1}{N}
\sum_b K^{(2)}_b \delta\bigg(X-\frac{f_b T^2}{2m}\bigg),
\nonumber\\
&&
~K^{(2)}_b 
=
\frac{\sqrt{2}\gamma \sqrt{m\pi}}{\sqrt{\Delta}\hbar}
\sum_{a} \frac{C^*_{a}(H_O )_{ab}
(H_O )_{ba}C_a }{|f_a -f_b |\sqrt{|f_a |}}.
\label{J^(2)_{1;a=a'}}
\eee
\v
\n
\u{The case $a\neq a'$}

\v
\n
In this case, $s'(X)\neq s(X)$ so $\Phi_{a'a;b}(X,s(X),s'(X))$ does not vanish and is 
a function of $X$ of $O(N)$, 
producing a stationary phase. We need to pin down the position. 
(We write $s_{ba}(X)=s(X)$ and $s'_{bab}(X)=s'(X)$.)
By the stationarity in $s$ and $s'$, eq.(\ref{s(X)zero}) and a
 similar equation 
for $s'$ with $a$ replaced by $a'$ hold;
\be
X-\xi_{ba'}(T,s'(X))
=-\frac{f_{a'}T^2}{2m} s'(X)^2 . 
\label{s'(X)zero}
\ee
Eqs.(\ref{s(X)zero}) and (\ref{s'(X)zero}) assure that
 we can limit our discussions near $s(X)=0$ and $s'(X)=0$,
 as stated just below (\ref{s(X)zero}). 

Now, using the stationarity in $s$ and $s'$,  
the stationary condition
 of $\Phi_{a'a;b}(X,s(X),s'(X))$ with respect to $X$ can be written as, 
\bea*
&&\hspace{-0.3cm}
0=\frac{d\Phi_{a'a;b}(X,s(X),s'(X))}{dX}
=\frac{\partial \Phi_{a'a;b}(X,s(X),s'(X))}{\partial X}
\\&&\hspace{-0.4cm}
=(f_a s(X) +f_b (T-s(X))-((f_{a'} s'(X) +f_b (T-s'(X))) 
\\&&\hspace{-0.4cm}
=(f_a -f_b )s(X)-(f_{a'}-f_b )s'(X).
\eea*
Higher derivatives are obtained by differntiating the
 stationary equation 
of $s$ or $s'$.  Differentiate (\ref{omegas}) 
 by $X$;
\bea*
&&
1-2\frac{1}{2m}(f_a -f_b )\frac{ds}{dX} T+\frac{1}{2m}(2f_a -f_b ) 
2s\frac{ds}{dX}=0 ,
\\&&
-2\frac{1}{2m}(f_a -f_b )\frac{d^2 s}{dX^2} T
\\&&
~~~~~~~+\frac{1}{2m}(2f_a -f_b ) 
2\left\{s\frac{d^2 s}{dX^2}+\left(\frac{d s}{dX}\right)^2 \right\}=0 .
\eea*
Setting $s=0$ in the first eqution, 
$(f_a -f_b )(ds/dX)=m/T$,  
which is inserted into the second. Thus one gets 
\bea*
&&
(f_a -f_b )\frac{d^2 s}{dX^2}=\frac{m^2}{T^3}\frac{(2f_a -f_b )}{(f_a -f_b )^2}.
\eea*
Using the similar equations for $s'$, with the replacement $a\rightarrow a'$, 
we get at $s(X)=0$,~$s'(X)=0$, 
\bbb
&&
\frac{d\Phi_{a'a;b}}{dX}=0,~~~~~~~~~~~~\frac{d^2 \Phi_{a'a;b}}{dX^2}=0,
\nonumber\\&&
\frac{1}{3!}\frac{d^3 \Phi_{a'a;b}}{dX^3}=\frac{1}{3!}
(f_a -f_b )\frac{d^2 s}{dX^2}-(f_{a'} -f_b )\frac{d^2 s'}{dX^2}
\nonumber\\&&
\nonumber\\&&
\hspace{-0.9cm}=
\frac{m^2}{6T^3}
\frac{(f_a -f_{a'} )(f_a f_b +f_{a'}f_b -2f_a f_{a'}) }
{(f_a -f_b )^2 (f_{a'} -f_b )^2}
\equiv \frac{C_{a'ab}}{T^3} .
\label{Ca'ab}
\eee
In this way, $\Phi_{a'a;b}$ becomes 
\be
\Phi_{a'a;b}
~\sim~
\frac{\ri N}{\hbar}\frac{C_{a'ab}}{T^3}\bigg(X-\frac{f_b T^2}{2m}\bigg)^3 .
\label{33}
\ee
Here we use the formula of symmetric integration
\bea*
&&
\lim_{N\rightarrow\infty}\exp\ri \alpha N(X-a)^3 =\lim_{N\rightarrow\infty}\cos \alpha 
N(X-a)^3 
\\&&
=(|\alpha|N)^{-1/3}\gamma'\delta(X-a),
~~~~~~~\gamma'=\int_{-\infty}^{\infty} dz\cos^3 z .
\eea*
Just as in the discussions for $J^{(1)}$ given in (\ref{1}) and (\ref{2}), 
the factors which are independent of
 $N$ become unity,
 since $N\rightarrow \infty$ is taken before $\Delta\rightarrow 0$.
The result for $a\neq a'$ is thus obtained;
\bbb
&&
J^{(2)}_{1;a\neq a'}=
\sum_{a\neq a';b'}(-\ri/\hbar)(\ri/\hbar)C^*_{a'}(H_O )_{a'b}
(H_O )_{ba}C_a 
\nonumber\\&&
\times \sqrt{\frac{1}{\pi\Delta^2}}\frac{Nm}{2\pi\hbar T}\frac{2\pi}{\sqrt{DD^*}}
\nonumber\\&&
\times\frac{1}{N}\frac{\pi\hbar 2m}{|f_a -f_b | |f_{a'} -f_b |T}
\frac{\gamma'\hbar^{1/3}T}{C^{1/3}_{aa'b}N^{1/3}}
\delta\bigg(X-\frac{f_b T^2}{2m}\bigg)
\nonumber\\&&
~~~~~~~~~=\frac{1}{NN^{1/3}}\sum_b K'^{(2)}_b \delta\bigg(X-\frac{f_b T^2}{2m}\bigg),
\nonumber
\\&&
\hspace{-0.7cm}
K'^{(2)}_b =
\frac{2\gamma' m \sqrt{\pi}}{\Delta\hbar }
\sum_{a\neq a'} {\rm Re}\frac{C^*_{a'}(H_O )_{a'b}(H_O )_{ba}C_a 
\hbar^{1/3}}{|f_a -f_b ||f_{a'}-f_b | C_{a'ab}^{1/3}}.
\label{J^(2)_{1;aneqa'}}
\eee
 Here ${\rm Re}$ signifies the real part. Under the 
exchange $a\leftrightarrow a'$, both 
$C^*_{a'}(H_O )_{a'b}(H_O )_{ba}C_a $ and $C_{a'ab}^{1/3}$ become complex 
conjugate. So the summation over all $a\neq a'$ is equivalent to apply 
${\rm Re}$. The results (\ref{J^(2)_{1;a=a'}})
 and (\ref{J^(2)_{1;aneqa'}}) have been checked by
interating by $X$ first and then by $s,~s'$.

\subsection{\label{02} Calculation of $J^{(2)}_2 $}

The remaining term of $O(([H_O ]^{\rm nd})^2 )$ appearing in $J^{(2)}$ is 
$\Psi^{(0)*}(b,X,T)\Psi^{(2)}(b,X,T)+$c.c., see (\ref{2nd2}).
 Actually, it is not necessary to 
calculate this term from the start. One can invoke to the 
normalization condition (\ref{normal}) with $k=2$. 
 Consider the result of $J^{(2)}_{1}$.
 If we write $J^{(2)}_1 =
\sum_b J^{(2)b}_1$, then 
\bbb
&&
J^{(2)b}_1 (X,T)=
\Psi^{(1)*}(b,X,T)\Psi^{(1)}(b,X,T)
\nonumber\\
&&
\hspace{-0.4cm}
=
\sum_{a,a'}V_{a'a;b}
C^*_{a'} (H_O )_{a'b}(H_O )_{ba}C_a \delta(X-\xi_b (T)).
\label{V}
\eee
$V_{a'a;b}$ is given by the sum of
 (\ref{J^(2)_{1;a=a'}}) and (\ref{J^(2)_{1;aneqa'}}) and 
has different forms for $a=a'$ and $a\neq a'$. 
As dictated from the definition of $U^{(2)}$ of (\ref{U2}), 
we can write $J^{(2)}_2 =\sum_b J^{(2)b}_2 $, where $J^{(2)b}_2 $ has the form   
\bbb
&&
J^{(2)b}_2 (X,T)=\Psi^{(0)*}(b,X,T)\Psi^{(2)}(b,X,T)+{\rm c.c.}
\nonumber\\&&\hspace{-0.6cm}
=
\sum_{c,a} W_{ba;c}C^*_b (H_O )_{bc}(H_O )_{ca}C_a \delta(X-\xi_b (T)).
\label{W}
\eee
Note the differnce of the index structure of~$b$ between (\ref{V}) and (\ref{W}). 
 Inserting these two into (\ref{normal}) ($k=2$),  
\bea*
&&
\sum_{c,a,b} W_{ba;c}C^*_b (H_O )_{bc}(H_O )_{ca}C_a 
\\&&~~~~~
=-
\sum_{a,a',b}V_{a'a;b}
C^*_{a'} (H_O )_{a'b}(H_O )_{ba}C_a 
\eea*
is obtained. After renaming the index of $W$, and recalling that the above equation holds for any $C_a$ and any operator $H_O$,  
 $W_{a'a;b}=-V_{a',a;b}$ follows.  
 Thus we arrive at for $a=a'$ and $a\neq a'$ separately; 
\bbb
&&
\hspace{-0.6cm}J^{(2)}_{2;a=a'}(X,T)
=-\frac{1}{N} 
\frac{\sqrt{2}\gamma\sqrt{m\pi}}{\sqrt{\Delta}\hbar}
\nonumber\\&&
\times\sum_{a,b} 
\frac{C^*_{a} (H_O )_{ab}(H_O )_{ba}C_a }
{|f_a -f_b |\sqrt{f_a}}\delta(X-\xi_{a} (T)),
\label{J^(2)_{2;a=a'}}
\\&&
\hspace{-0.6cm}J^{(2)}_{2;a\neq a'}(X,T)
=-\frac{1}{NN^{1/3}} 
\frac{2\gamma'm \sqrt{\pi}}{\Delta \hbar}
\label{J^(2)_{2;aneqa'}}
\\
&&\hspace{-0.7cm}
\times\sum_{a\neq a',b} 2{\rm Re}
\frac{C^*_{a'} (H_O )_{a'b}(H_O )_{ba}C_a \hbar^{1/3}}
{|f_{a'} -f_b ||f_a -f_b |C_{a'ab}^{1/3}}
\delta(X-\xi_{a'} (T)).
\nonumber
\eee

\subsection{\label{final}~The result for $J(X,T)$ up to $([H_O ]^{\rm nd})^2$}

Collecting all the results of (\ref{ideal}), (\ref{J^(1)}), 
(\ref{J^(2)_{1;a=a'}}),(\ref{J^(2)_{1;aneqa'}}) 
(\ref{J^(2)_{2;a=a'}}) and (\ref{J^(2)_{2;aneqa'}}), 
the signal function up to $O([H_O ]^{\rm nd})^2$ is 
\bbb
J(X,T)&=&J^{(0)}(X,T)+J^{(1)}(X,T)
+J^{(2)}_{1;a=a'}(X,T)
\nonumber\\&&\hspace{-0.9cm}
+J^{(2)}_{1;a\neq a'}(X,T)
+J^{(2)}_{2;a=a'}(X,T)+J^{(2)}_{2;a\neq a'}(X,T)
\nonumber\\
&&
\hspace{-1.8cm}=
\sum_b \bigg[|C_b |^2 +\frac{A_b}{N} 
+\frac{B_b}{NN^{1/3}}\bigg]
\delta(X-f_b T^2/2m).
\label{AbBb}
\eee
Here each coefficient is given by
\bbb
&&
A_b =A_{b1}+A_{b2}, 
\nonumber\\&&
A_{b1}=
\frac{\sqrt{\pi}}{\Delta}
\sum_{a}\frac{2{\rm Im}~\!C^*_b (H_O )_{ba}C_a}{|f_a -f_b |},
\label{Ab1}
\\
&&
A_{b2}=
\frac{\sqrt{2}\gamma\sqrt{m\pi}}{\sqrt{\Delta}\hbar}
\bigg(\sum_{a} \frac{C^*_{a}(H_O )_{ab}
(H_O )_{ba}C_a }{|f_a -f_b |\sqrt{|f_a |}}
\nonumber\\&&~~~~~~~~~~~~~~~~~~~-\sum_{c} 
\frac{C^*_{b} (H_O )_{cb}(H_O )_{ca}C_b }
{|f_b -f_c |\sqrt{f_a}}\bigg),
\label{Ab2}
\\&&
B_b =
\frac{2\gamma'm \sqrt{\pi}\hbar^{1/3}}{\Delta \hbar} 
\nonumber
\\&&
~~~\times{\rm Re}\bigg(
\sum_{a\neq a'}\frac{C^*_{a'}(H_O )_{a'b}(H_O )_{ba}C_a }
{|f_a -f_b ||f_{a'}-f_b | C_{a'ab}^{1/3}}
\nonumber
\\&&~~~~~~~~~~~~~
-\sum_{b\neq a,c} 
\frac{C^*_{b} (H_O )_{bc}(H_O )_{ca}C_a }
{|f_{b} -f_c ||f_a -f_c |C_{bac}^{1/3}}\bigg).
\label{Bb}
\eee
The origin of the power of $1/N$ is the 
stationary phase integrations by $s$, $s$' and $X$.
 Each $s$-integration brings us $1/\sqrt{N}$,
 but the results of the $X$-integration (or equivalently of the formula (\ref{formula}))
 depend on the situation; $1/\sqrt{N}$ for $A_{b1}$, $(1/\sqrt{N})^0$ for $A_{b2}$ (no
stationary phase) and for $B_b$, $1/N^{1/3}$.
 In Appendix \ref{Neglect}, 
 the order of neglected terms are estimated. They are shown to be down by 
at least $1/\sqrt{N}$ compared with (\ref{AbBb}).  

As $T\rightarrow 0$, above $J(X,T)$ goes over 
to $J(X,0)=\delta(X)$ by the same relation that follows from 
the normalization condition.
Note further that when $[H_O ,O]=0$, all the correction terms
 are absent and the ideal result becomes exact up to the order considered here.  

\subsection{\label{Tindep}T-independence of $A_{b1,b2}$ and $B_b$}

Above correction terms $A_{b1}$, $A_{b2}$ and $B_b$ are independent of $T$, 
besides $|C_b |^2$. The reason is simple;
 as we have seen above, 
the fluctuation into any channel $b$ occurs at $t=0$ 
for $N\rightarrow\infty$,
 and after that the state develops by $U_{N,b}$ of (\ref{Uas}) and (\ref{fa}), 
so the claasical time evolution is realized.
 Consider the signal function
 $J(b,X,T)=|\Psi(b,X,T)|^2$ defined for each channel,
 i.e. one term in the sum (\ref{J}) of $J(X,T)=\sum_b J(b,X,T)$.  Then 
once the above fluctuation at $t=0$ is taken into account, $J(b,X,T)$   
 evolves as $\delta(X-\xi_b (T))$ without any further $T$ dependence.
 Therefore, when we integrate $J(b,X,T)$ 
in $R_b$ of (\ref{intR}) and define 
\[
J(b,T)\equiv \int_{R_b}J(b,X,T)~=~\int_{R_b}J(X,T)dX,
\] 
then $J(b,T)$ is independent of $T$. 
 Since this holds for any $H_O$ or for 
any parameters in the thory, we conclude that 
$A_{b1},~A_{b2}$ or $B_b$ does not depend on $T$. This statement 
applies even if one includes higher order of $H_O$. However, when 
the diffusion processe is taken into account,
 the above assertion does not hold. 

\subsection{\label{numerical}Numerical estimates}

Here we try to estaimate the order of numerical values of the correction terms. 
 Consider first $A_{b1}$ given in (\ref{Ab1}); 
The numerator is the order of enegy of the object.
 As discussed in deriving (\ref{10-12}), $f_a a$  
order of the energy a partiacle in the detector recerives from the object. Thus we regard $C^*_b (H_O )_{ba}C_a$ is  
is of the same oredr as $|f_a -f_b |a \sim f_a a$. 
In this way, adopting 
 $\Delta =10^{-3}$cm, $a=10^{-7}$cm, 
\be
\frac{A_{b1}}{N}\sim\frac{1}{N}\times\frac{a}{\Delta}\sim \frac{1}{N}
\times 10^{-4}.
\label{firstcorr}
\ee
The condition $a/N\Delta<\!\!<1$ is our requirement for the measuring 
process to be sensible. Thus one has to require $N>\!\!>10^{4}$. We have stressed thoughout the paper that 
the limit $N\rightarrow\infty$ is kaken before $\Delta\rightarrow 0$, which is 
denoted by $\Rightarrow$. The requirement $a/N\Delta<\!\!<1$ agrees with this limit
 and at the same time gives the precise condition of this limit. In our case, the ``macroscopic'' number 
of particles in the detector required to produce experimental 
signal is expeessed as $N>\!\!>10^{4}$, which can be 
much smaller than, for example, the Avogadro number. The last conclusion is 
stated frequently for realistic devices. 

Next, we estimate $A_{b2}$ of (\ref{Ab2}). Similarly as above, the factor 
$C^*_{b} (H_O )_{cb}(H_O )_{ca}C_b$ is regarded as the same
 order with $|f_a -f_b |f_a a^2 \sim f_a^2 a^2$.  
Then apart from the numerical constant, $A_{b2}\sim ( a^2 /\hbar) \sqrt{|f_a |m/\Delta}$. In this way,   
\bbb
&&
\hspace{-0.8cm}\frac{A_{b2}}{N}\sim\frac{1}{N}\sqrt{\frac{|f_a |a}{\hbar^2 /(ma^2 )}}
\bigg(\frac{a}{\Delta}\bigg)^{1/2}\sim \frac{1}{N}\times 4.0\times 10^{-2}.
\label{secondcorr}
\eee
Here we have used the same values as in (\ref{10-12}). For the electron, 
 $\hbar^2 /(ma^2 ) 
\sim 10^{-13}{\rm erg}$. 
The above (\ref{secondcorr}) is much larger than (\ref{firstcorr}) and may be the 
main correction to the ideal zeroth order result of (\ref{AbBb}). 
The reason why this term is large is, as stated before, understood if we look at (\ref{Phia'a;bb}).  When $a=a'$,  
$\omega_{aa;b}(X,s_{ab}(X),s'_{ab}(X))=0$; after inserting the stationary condition of $s$ and $s'$, 
 the exponent of $\Phi$ becomes $O(1)$, not $O(N)$, so the cancelation of the phase as a function of $X$ does not 
occur and the resulting value becomes large.  
When, for example, $N=10^{5}$ particles are participating in giving the signal, 
 the corrction of the order $10^{-7}$ is expected. This can be within the experimental 
confirmation.   

Finally as for $B_b$, 
we note $C_{bac}/\hbar \sim m^2 /f_a$ by (\ref{Ca'ab}).  
 In this way, we get  
\bea*
&&
\frac{B_b}{NN^{1/3}}=
\frac{1}{NN^{1/3}}\frac{ma^2}{\hbar }
\times \frac{1}{(C_{abc}/\hbar)^{1/3}}\times 
\frac{a}{\Delta}
\\
&&=\frac{1}{NN^{1/3}}\bigg(\frac{f_a a}{\hbar^2 /(ma^2 )}\bigg)^{1/3}\frac{a}{\Delta}
\sim\frac{1}{NN^{1/3}}\times 2.5\times 10^{-4}.
\eea*
This is a small number.

\section{\label{Discussions}Discussions}

In order to see whether or not our result (\ref{AbBb}) is specific to our model, 
 let us consider the general Hamiltonian (\ref{HODI}),
 combined with (\ref{general})
 and apply the arguments of Appendix \ref{nonsepa}. 
When $N\rightarrow\infty$, the leading term agrees with the ideal result 
 of (\ref{AbBb}). This is because $H_O$ is negligible and the 
object stays in the prescribed eigen-state $|b>$ of $O$.
 For fixed $b$, the detector evolves 
by $H_D +H_I (\lambda_b ,[x_i ])$,
 so the stationary path $X_b^{\rm st}(t)$ depends on 
$b$.  By normalization, the signal function (\ref{defJ})
 applied at $t$ takes the form  
 \[
J(X,t)=\sum_b |C_b |^2 \delta (X-X_b^{\rm st}(t))
\]
for general case also. Expanding in $[H_O ]^{\rm nd}$, we use (\ref{U_O}), 
with one $s$ integration accompanying each $H_O$.
  Since each $s$ integral is dominated by the 
 stationary phase, one $s$ brings us $1/\sqrt{N}$, 
irrespective of the form of $h$ and $g$.
 We conclude that $k$-th order correction terms 
containts at least the factor $(1/\sqrt{N})^k$. 
Whether extra 
factors of $1/N$ appears or not depends on the form of 
 $H_D +H_I (\lambda_b ,[x_i ])$, 
 just as in our model extra factors of $1/N$ 
appeared through $X$-interation. This was pointed out
 just below (\ref{Bb}). 
We have to study more realistic detection process and 
fix the above correction terms, including the numerical estimates. 
In doing so, some simplification of the Hamiltonian will be required of course 
to make the problem tractable.   

Our scheme can also be applied directly to the system described by the quantized 
field. Indeed, the field theory is much more suited to handle 
the macroscopic system, especially when taking the thermodynamic limit. 
Also, the measurement theory in the relativistic case can be 
 studied using the field theory, since the formalism of the 
relativistic field theory is firmly established.
\\~~
Finally, the most difficult problem of reduction, or ein-selection,
 is left out of the discussions in this paper. 
The extension of the dynamical reduction thory \cite{Rimini}  
to the relativistic case is controversial. 
However, in our case the application of the stationary phase to
the relativistic field theory is straightforward. 
Also our results in this paper are independent of the precise
 mechanism of the reduction process, since 
we rely solely on the Schr$\ddot{\rm o}$dinger equation,
 which holds independently of how the reduction is realized.  

\appendix
\section{\label{nonsepa}Non-separable case}

To obtain the equation of motion for a macrovariable
 is the same problem of how to get the effective thory of 
a collective mode in many particle system.
 There are several methods but here we select the one which consists  
 of inserting a delta-function in the path-integral.
 In the limit $N\rightarrow\infty$, 
it becomes equal to the method of the Legendre transformation \cite{fukudaa}.
We consider the case where $X$ is given in collective notation $[x_i ]$ as 
$g([x_i ])/N$. Extension to the case where $g$ includes $[p_i ]$ is not difficult. 
  
\vspace{0.1cm}
\n
\u{The signal function}

\vspace{0.1cm}
\n
Let us write the wave function of the detector plus 
object system at some fixed time as $\Psi([x_i ] ,x)$.
 For any thermodynamically normal macroscopic system, 
the wave function has the form \cite{FGKX}
\be
\Psi([x_i ] ,x)=G~\!{\rm e}^{-F},
\label{FG}
\ee
where $F=F([x_i ] ,x)$ is an extensive quantity and is of the order $O(N)$, so 
we write it as $F([x_i ],x)=N{\mathcal F}([x_i ],x)$. On the other hand,
  $G=G([x_i ],x)$ describes the microscopic details. By normalizability, 
the real part of ${\mathcal F}([x_i ] ,x)$ is positive
 definite and when ${\mathcal F}([x_i ],x)$ changes by a finite amount,  
it describes macroscopically different state of the system.
 When we sum up wave functions 
having different values of ${\mathcal F}$, it represents the mixed state. The pure state 
is obtained by summing up various $G$'s, with the same ${\mathcal F}$. 
  The signal function defined in (\ref{J}) 
 can be generalized to the 
 non-separable case as follows.
\bbb
&&
J(X)=N\int dx \int \prod_i dx_i 
\nonumber\\&&
~~~~\times|\Psi([x_i ] ,x)|^2 
\delta(NX-g([x_i ] ))
\label{defJ}
\\
&&~~~~
=N\int dx\int \prod_i dx_i \int_{-\infty}^{\infty} dj 
\nonumber\\&&
~~~~\times\exp \bigg(-N2{\rm Re}{\mathcal F}([x_i ],x)
+\ri \{NX-g([x_i ])\}j\bigg)
\nonumber\\
&&
~~~~\equiv  \int_{-\infty}^{\infty} dj \exp (-H(j)+\ri NjX)\equiv{\rm e}^{-K(X)}.
\label{KX}
\eee
We have dropped microscopic $G$ term since it does not affect the stationary 
condition for $X$. Note here that $J(X)$ is real by definition, so 
$K(X)$ is a real quantity and positive definite. Now $H(j)$ is extensive, so 
 the function $H(j)$ and $K(X)$ are both extensive proportional to 
$N$.
 The results (\ref{FG}) and (\ref{KX}) have been obtained \cite{FGKX} 
 by adoping the field theory 
to describe the macroscopic system, which automatically takes
 into account the statistical factor $1/N!$. 
The essential condition for these two equations to hold is the short 
range character of the interation among constituent particles 
in the macroscopic system. Although 
the field theoretical approach is not taken in this paper, 
here we apply (\ref{FG}) and (\ref{KX}) 
to the quantum mechanical $N$-particle system,  
since these results express the general property of the thermodynamically 
normal system. 
Writing $K(X)=N{\mathcal K}(X)$, we first expand ${\mathcal K}(X)$ around the 
the stationarity solution $X=X^{\rm st}$ satisfying
 $\partial {\mathcal K}(X)/\partial X=0$.
Suppose the second derivative at the stationary point is positive;
 ${\mathcal K}^{(2)}(X^{\rm st})>0$. 
(Since ${\mathcal K}(X)$ is positive definite,
 at least one stationary point with ${\mathcal K}^{(2)}>0$ exists. 
The solution with ${\mathcal K}^{(2)}<0$ represents an unstable state.)
In the limit $N\rightarrow\infty$, the analog of (\ref{formula}) for 
the case of positive definite ${\mathcal K}(X)$ holds; 
\bea*
&&
\lim_{N\rightarrow \infty}{\rm e}^{-N{\mathcal K}(X)}
=
{\rm e}^{-N{\mathcal K}(X^{\rm st})}
\sqrt{\frac{2\pi}{-N{\mathcal K}''(X^{\rm st})}}
\delta(X-X^{\rm st}).
\eea*
This is equal to $\delta(X-X^{\rm st})$ by normalization. 

When $J(X)$ has the distribution in $X$,
 it represents the mixed state. Indeed, 
 the signal function is rewritten as 
\bea*
&&
J(X)=\int dY J(Y) \delta(X-Y)
\\&&\hspace{-0.6cm}
=\lim_{N\rightarrow\infty}\int dYJ(Y) 
\sqrt{N/\pi}\exp(-N(X-Y)^2 ).
\eea*
In this form, $J(X)$ of (\ref{gauss}) is seen
 to be a superposition of different ${\mathcal F}$'s,
 with the weight $J(Y)$.  
 In real detection process, the detector is so arranged that 
${\mathcal F}$ depends on the eigen-value $\lambda_a$ of the object operator $O$
 to be measured. Writing the time $t$ 
explicitly, 
$F(\lambda_a ,[x_i ],t)=N{\mathcal F}(\lambda_a ,[x_i ],t)$ and 
 $K=N{\mathcal K}$ in (\ref{KX}) becomes 
$N{\mathcal K}(\lambda_a ,X,t)$.  Then the stationary solution
 of $\partial {\mathcal K}/\partial X=0$ 
depends on $t$ and $a$; $X=X^{\rm st}_a (t)$.
For the model of (\ref{fa}), this is written as $\xi_a (T)$ in (\ref{xia}). 
 
The equation of motion of $X^{\rm st}(t)$ itself 
 can be obtained by the method of double paths Legendre transform, as 
discussed in detail in \cite{fukudaa}.  

\section{\label{Phibaaa}Calcultion of $\Phi_{ba}$}

By (\ref{1st}), (\ref{psi^0}) and (\ref{psi^1}), $J^{(1)}(X,T)$ 
 is given by 
\bbb
&&
J^{(1)}(X,T)=\sum_b \Psi^{(0)*} (b,X,T)\Psi^{(1)}(b,X,T)+{\rm c.c.}
\nonumber\\
&&
=\left(\frac{1}{\pi\Delta^2 \rho}\right)^{1/2}(-\ri/\hbar)\sum_{ab}C^*_b (H_O )_{ba}C_a 
\nonumber\\&&
\times \int_0^T ds\times\exp\ri\{(\theta_a -\theta_b )s\}\exp{\Phi}_{ba}+{\rm c.c.}. 
\label{Phiba}
\eee
By $X-\xi_b (T)=R_{ba}(X,T,0)$, $\Phi_{ba}$ 
is expressed by 
\bbb
&&
\Phi_{ba}=-\frac{N^2 m^2 }{2\hbar^2 T^2 }
\left(\frac{R_{ba}(X,T,s)^2}{D}+\frac{R_{ba}(X,T,0)^2}{D^*}\right)
\nonumber\\
&&
~~~~~~~~~~~~~+
\frac{Nm\ri}{2\hbar T}\{R_{ba}(X,T,s)^2 -R_{ba}(X,T,0)^2 \}
\nonumber
\\&&
~~~~~~~~~~~~~+\frac{\ri}{\hbar}
N\{XQ_{ba}(T,s)+P_{ba}(T,s)
\label{Phibaa}
\\&&
~~~~~~~~~~~~~~~~~~-XQ_{ba}(T,0)-P_{ba}(T,0)\}.
\nonumber
\eee
Here we use (\ref{DD*}) and rewrite the first term of $\Phi_{ba}$ as 
\bbb
&&
\frac{N^2 m^2 }{2\hbar^2 T^2 }\left(\frac{R_{ba}(X,T,s)^2}
{D}+\frac{R_{ba}(X,T,0)^2}{D^*}\right)
\nonumber\\&&
=\frac{1}{\rho\Delta^2}\bigg(X-\displaystyle \frac{\xi_{ba}(T,s) 
+\xi_{ba}(T,0)}{2}\bigg)^2 
\nonumber
\\
&&
+\frac{1}{4\Delta^2 \rho}(\xi_{ba}(T,s)-\xi_{ba}(T,0))^2 
\label{Phiba'}
\\&&
+\frac{\ri Nm}{2\rho\hbar T}
 \{(X-\xi_{ba}(T,s))^2 -(X-\xi_{ba}(T,0))^2 \}.
\nonumber
\eee
Approximating $\rho=1$, 
 $\Phi_{ba}$ is now rewritten as
\bbb
&&
\Phi_{ba}=
-\frac{1}{\Delta^2}\bigg(X-\displaystyle 
\frac{\xi_{ba}(T,s) +\xi_{ba}(T,0)}{2}\bigg)^2 
\nonumber\\&&
\hspace{-0.9cm}-\frac{1}{4\Delta^2 }(\xi_{ba}(T,s)-\xi_{ba}(T,0))^2 
+\frac{\ri N}{\hbar}\omega_{ba}(X,T,s),
\label{Phiccc}
\eee
with $\omega_{ba}(X,T,s)$ defined by 
\bbb
&&\hspace{-0.5cm}
\omega_{ba}(X,T,s)=X(Q(T,s)-Q(T,0))+P(T,s)-P(T,0)
\nonumber\\&&
=X(f_a -f_b )s 
-\frac{f_a^2}{6m}s^2 (3T-2s)
\nonumber\\&&
~~-\frac{f_b^2}{6m}\{(T-s)^3 -T^3 \} -\frac{f_a f_b}{2m}s(T-s)^2 .
\label{omegaaa}
\eee

\section{\label{Phiabc} Calculation of $\Phi_{a'a;b}$}

By (\ref{psi^1}), the explicit form of $\Phi_{a'a;b}$ defined in (\ref{Phia'ab}) is 
\bbb
&&
\Phi_{a'a;b}=-\frac{N^2 m^2 }{2\hbar^2 T^2 }\left(\frac{R_{ba}(T,s)^2}{D}+\frac{R_{ba'}(T,s')^2}{D^*}\right)
\nonumber\\&&
~~~~~~~~~~+
\frac{Nm\ri}{2\hbar T}\{R_{ba}(T,s)^2 -R_{ba'}(T,s')^2 \}
\nonumber\\&&
~~~~~~~~~~+\frac{\ri}{\hbar}N\{XQ_{ab}(T,s)+P_{ab}(T,s)
\nonumber\\&&
~~~~~~~~~~-XQ_{a'b}(T,s')-P_{a'b}(T,s')\}.
\label{Phissapp}
\eee
As in (\ref{Phiba'}), we rewrite
\bbb
&&
\frac{N^2 m^2 }{2\hbar^2 T^2 }\left(\frac{R(T,s)^2}{D}
+\frac{R(T,s')^2}{D^*}\right)
\nonumber\\&&
=\frac{1}{\rho\Delta^2}\bigg\{\bigg(X-\displaystyle \frac{\xi_{ba}(T,s) +\xi_{ba'}(T,s')}{2}\bigg)^2 \bigg\}
\nonumber\\&&
~~~+\frac{1}{4\Delta^2 \rho}(\xi_{ba}(T,s)-\xi_{ba'}(T,s'))^2 
\nonumber\\&&
~~~-\frac{\ri Nm}{2\rho\hbar T} \{(X-\xi_{ba}(T,s))^2 -(X-\xi_{ba'}(T,s'))^2 \}.
\nonumber
\eee
By setting $\rho=1$, $\Phi_{a'a;b}$ is written as 
\bbb
&&
\Phi_{a'a;b}=
-\frac{1}{\Delta^2}
\bigg(X-\displaystyle \frac{\xi_{ba}(T,s) +\xi_{ba'}(T,s')}{2}\bigg)^2 
\nonumber
\\&&
-\frac{1}{4\Delta^2 }(\xi_{ba}(T,s)-\xi_{ba'}(T,s'))^2 
+\frac{\ri N}{\hbar}\omega_{a'a;b}(T,s),
\label{Phia'a;b}
\\
&&
\omega_{a'a;b}(T,s)=\omega_{ba}(T,s)-\omega_{ba'}(T,s')
\label{abc}\\&&=
X(Q_{ba}(T,s)-Q_{ba'}(T,s'))+P_{ba}(T,s)-P_{ba'}(T,s').
\nonumber
\eee

\section{\label{integX}Integration by $X$}

Since each term in $\sum_b$ of (\ref{J^(11)}) is given by 
 $K_b^{(1)}\delta(X-\xi_b (T))$, $K_b^{(1)}$ is calculable by 
integrating over $X$ first and then by $s$ for fixed $b$. 
After integration by $X$, the term 
\bea*
&&\hspace{-0.5cm}
\exp\left[
\frac{-1}{\Delta^2 \rho}\left\{
X-\frac{\xi_{ba}(T,s)+\xi_{ba}(T,0)}{2}\right\}^2 
+\frac{\ri N}{\hbar}X(f_a -f_b )s\right]
\eea*
changes into 
\bea*
&&
\sqrt{\pi\Delta^2 \rho}\exp\left[
-\frac{N^2 \Delta^2 \rho}{4\hbar^2}(f_a -f_b )^2 s^2 
\right.
\\
&&\left.
~~~~+\frac{N\ri}{2\hbar}\bigg(\xi_{ba}(T,s)+\xi_{ba}(T,0)\bigg)(f_a -f_b )s\right].
\eea*
Then the dominant region of $s$-integration is $s=O(1/N)$. 
The second term in $[\cdots\cdots]$ can be shown 
to be $O(s^3 )$, if it is combined with 
 $P_{ba}(T,s)-P_{ba}(T,0)$. Indeed,  
\bea*
&&\hspace{-0.3cm}
(\xi_{ba}(T,s)+\xi_{ba}(T,0))(f_a -f_b )s+P_{ba}(T,s)-P_{ba}(T,0)
\\&&
~~~~\sim (f_a -f_b )(2f_a -f_b )s^3 /6m.
\eea*
for small $s$. 
By $Ns=y$, $s$-integration now becomes
\bea*
&&
\frac{1}{N}\int_0^{TN} dy\times\exp\ri\{(\theta_a -\theta_b )y/N\}
\sqrt{\pi\Delta^2 \rho}
\\&&
\times\exp\left[
-\frac{\Delta^2 \rho}{4\hbar^2}(f_a -f_b )^2 y^2 \right] +~O(1/N^2).
\eea*
It gives $\hbar\pi/(N|f_a -f_b |)(1+O(1/N))$.  Thus we get 
\bea*
&&
\hspace{-1.0cm}\int dX \Psi^{(0)*}_b (X,T)\Psi^{(1)}_b (X,T)
=\frac{-\ri\sqrt{\pi}\sum_{a}C^*_b (H_O )_{ba}C_a}{N\Delta|f_a -f_b |}. 
\eea*
Adding term with c.c., we get (\ref{J^(11)}) and (\ref{J^(1)}). 

\section{\label{Neglect}Neglected terms for $N\rightarrow\infty$}

First, let us restrict the arguments up to $O([H_O ]^{\rm nd})^2$   
 and consider the correction to 
 $|C_b |^2$, $A_b$ and $B_b$ of (\ref{AbBb}). 
\\
1.~$|C_b |^2$;~~As stated in Sec.\ref{psi0}, the correction comes from the 
diffusion. 
Including the diffusion,
  $\Delta$ 
 is replaced by 
 $\sqrt{\rho}\Delta$. 
Since $\sqrt{\rho}=1+O(1/N^2)$, 
$|C_b |^2$ changes into $|C_b |^2 (1+O(1/N^2 ))$. 
Thus the neglected terms are $O(1/N^2 )$.
 The fact that it is small numerically has been  
checked in Sec.\ref{psi0}.  ~~
\\
2.~$A_b$;~~The correction comes from the higher order of the
 fluctuation around the stationary
 path, which is  down by $O(1/\sqrt{N})$ compared with the term retained. 
In obtaining $A_{b1}$ of (\ref{Ab1}),
 both $s$ and $X$ have fluctuations of  
 $O(1/\sqrt{N})$.
 As for $A_{2b}$, $s$ and $s'$ has the same size of the fluctuations but 
$X$ has no stationary point. 
 Therefore $A_{b1,2}$ are replaced by $A_{b1,2} (1+O(1/\sqrt{N})$.  
The leading order of neglected terms in $A_b /N$ is thus $O(1/N\sqrt{N})$. ~~
\\
3.~$B_b$;~~Besides $s$ and $s$' having the fluctuation of $O(1/\sqrt{N})$, 
 $X$ fluctuates near the stationary phase with the size of 
$O(1/N^{1/3})$, see (\ref{33}).
 Thus $B_b$ is replaced by $B_b (1+O(1/N^{1/3})$ and the corrections 
to $B_b /NN^{1/3}$ is $O(1/NN^{1/3})(1/N^{1/3})$. 
\\~~~
Next, consider the term $([H_O ]^{\rm nd})^k$ for $k\geq 3$. 
In the expansion (\ref{U_O}), one factor of $[H_O ]^{\rm nd}$ 
accompanies an integration over the parameter $s$. Integration 
around the stationary 
phase produces one factor $1/\sqrt{N}$, so $(1/\sqrt{N})^k$ appears. 
Extra factor of some power of $1/N$ may appear 
depending on the result of the $X$-intergation, as stated above. Therefore, 
we can say that $([H_O ]^{\rm nd})^k$ term has the power of 
at least $(1/\sqrt{N})^k$. 

Summarizeibg, eq.(\ref{AbBb}) is correct up to the order retained there, 
 the leading correction being $O(1/N\sqrt{N})$.

\end{document}